\documentclass[aps,prl,reprint,longbibliography,superscriptaddress]{revtex4-1}

\usepackage{setspace}
\usepackage{amsmath}
\usepackage{graphicx}
\usepackage[nearskip,margin = 0pt]{subfig}
\usepackage{subfig}

\usepackage{verbatim}
\usepackage{amsfonts}
\usepackage{amssymb}

\usepackage{epstopdf} 
\usepackage{xcolor}
\DeclareGraphicsExtensions{.pdf,.eps,.png,.jpg,.mps} 
\usepackage{hyperref}
\hypersetup{
    colorlinks=true,
    linkcolor=blue,
    citecolor=blue,
    filecolor=blue,      
    urlcolor=blue,
}

\begin{document}
\title{Forced Oscillatory Motion of Trapped Counter-Propagating Solitons}

\author{Chengying Bao}
\affiliation{T. J. Watson Laboratory of Applied Physics, California Institute of Technology, Pasadena, California 91125, USA}

\author{Boqiang Shen}
\affiliation{T. J. Watson Laboratory of Applied Physics, California Institute of Technology, Pasadena, California 91125, USA}

\author{Myoung-Gyun Suh}
\affiliation{T. J. Watson Laboratory of Applied Physics, California Institute of Technology, Pasadena, California 91125, USA}

\author{Heming Wang}
\affiliation{T. J. Watson Laboratory of Applied Physics, California Institute of Technology, Pasadena, California 91125, USA}

\author{Kemal \c{S}afak}
\affiliation{Cycle GmbH, Hamburg 22607, Germany}

\author{Anan Dai}
\affiliation{Cycle GmbH, Hamburg 22607, Germany}

\author{Andrey B. Matsko}
\affiliation{Jet Propulsion Laboratory, California Institute of Technology, Pasadena, California 91109, USA}

\author{Franz X. K$\rm\ddot{a}$rtner}
\affiliation{Center for Free-Electron Laser Science, Deutsches Elektronen-Synchrotron, Hamburg 22607, Germany}
\affiliation{Department of Physics and the Hamburg Center for Ultrafast Imaging, University of Hamburg, Hamburg 22761, German}

\author{Kerry J. Vahala}
\email{vahala@caltech.edu}
\affiliation{T. J. Watson Laboratory of Applied Physics, California Institute of Technology, Pasadena, California 91125, USA}

\begin{abstract}
Both the group velocity and phase velocity of two solitons can be synchronized by a Kerr-effect mediated interaction, causing what is known as soliton trapping. Trapping can occur when solitons travel through single-pass optical fibers or when circulating in optical resonators. Here, we demonstrate and theoretically explain a new manifestation of soliton trapping that occurs between counter-propagating solitons in microresonators. When counter-pumping a microresonator using slightly detuned pump frequencies and in the presence of backscattering, the group velocities of clockwise and counter-clockwise solitons undergo periodic modulation instead of being locked to a constant velocity. Upon emission from the microcavity, the solitons feature a relative oscillatory motion having an amplitude that can be larger than the soliton pulse width. This relative motion introduces a sideband fine structure into the optical spectrum of the counter-propagating solitons. Our results highlight the significance of coherent pumping in determining soliton dynamics within microresonators and add a new dimension to the physics of soliton trapping.
\end{abstract}

\maketitle

\begin{figure*}[t!]
\begin{centering}
\includegraphics[width=0.95\linewidth]{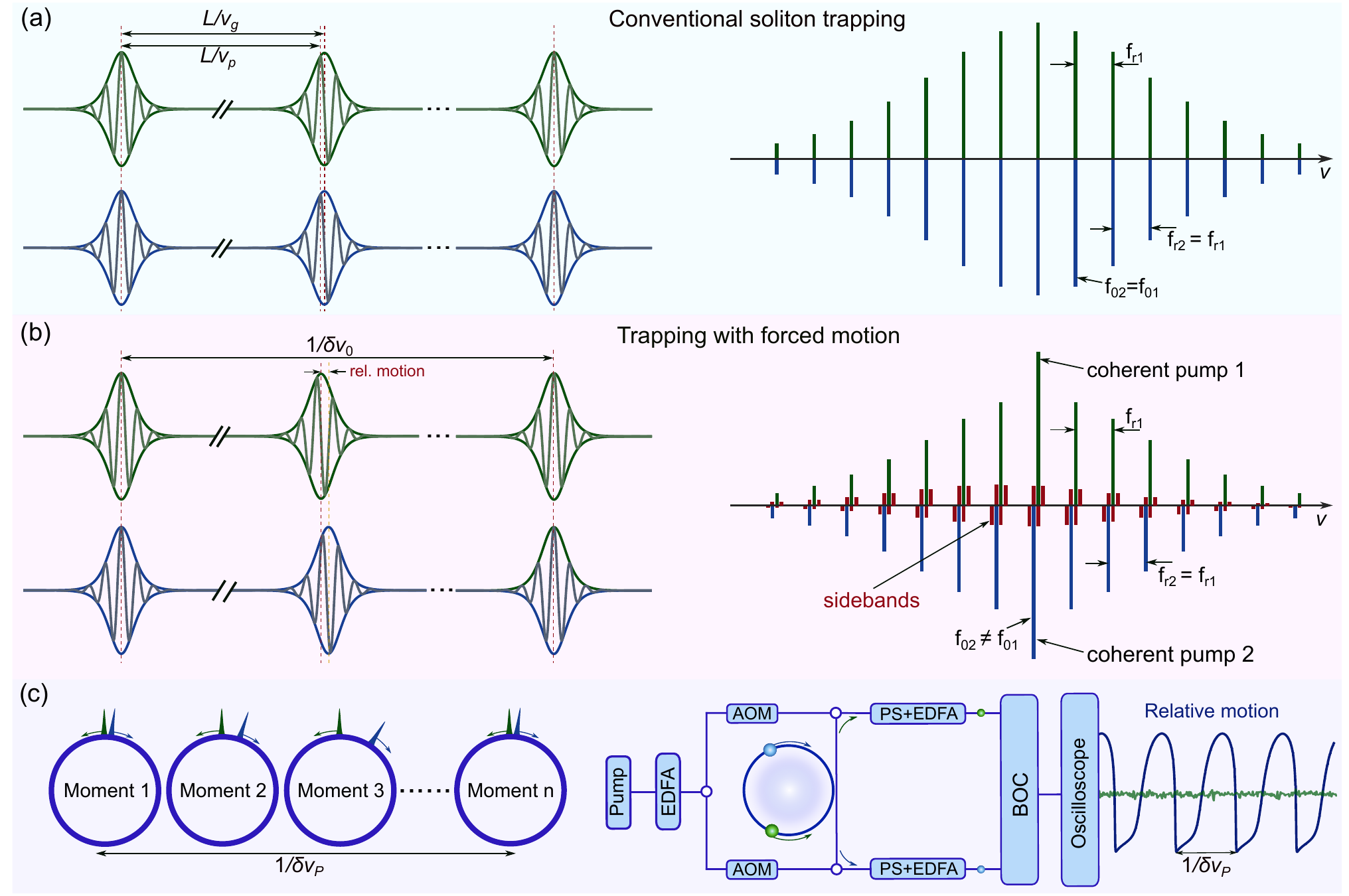}
\captionsetup{singlelinecheck=no, justification = RaggedRight}
\caption{{\bf Trapping with forced motion in counter-propagating solitons.} (a) In conventional soliton trapping without coherent pumps, both phase velocity and group velocity will be locked. In the frequency domain, both the repetition rate $f_r$ and carrier envelope offset frequency $f_0$ of the corresponding combs will be equal. Note that in the illustration the pulse envelopes are plotted to be exactly overlapped and the phase is identical, which are not necessary in experiments. (b) In coherently pumped systems, phase velocity and $f_0$ cannot be locked when using non-degenerate pumps. In this case, the group velocity of the solitons will be locked in an averaged way and there is relative motion in the trapping. This relative motion repeats with a period of 1/$\delta \nu_\text{P}$. In the frequency domain, the motion will induce sidebands around the main comb lines. (c) Illustration showing CP solitons at various moments in time. In the measurements, a balanced optical cross correlator (BOC) is used to measure relative soliton motion (blue line). The counter-pumping frequencies are controlled by two acousto-optical modulators (AOMs). The green line illustrates the expected BOC output signal for conventional (non oscillatory) trapping. EDFA: erbium-doped fiber amplifier, PS: pulse shaper.}
\label{Fig1Picture}
\end{centering}
\end{figure*}

The mutual interaction of two optical soliton pulses by way of the optical Kerr effect is known to impact their relative phase and group velocity leading to the phenomenon of optical trapping.  Solitons with close group velocities can be mutually trapped and travel at the same group velocity \cite{Menyuk_OL1987,Gordon_OL1989,Cundiff_PRL1999,Wise_NC2013,Vahala_NP2017Stokes,Gaeta_NP2018Synchronization}. Moreover, their pulse phase velocities can also be locked (i.e., are equal, see Fig. \ref{Fig1Picture}(a)) \cite{Cundiff_PRL1999} so that both the envelope and the carrier of the solitons travel at the same velocities. Recently, there has been considerable interest in a coherently pumped optical soliton \cite{Wabnitz_OL1993} that has been realized in mode locked fiber systems \cite{Coen_NP2010fiber} and high-Q microresonators \cite{Kippenberg_NP2014,Vahala_Optica2015,Weiner_OE2016,Gaeta_Ol2016Thermal}. Besides their practical importance for realization of compact microcomb systems \cite{Kippenberg_Science2018Review}, the coherent pumping of these solitons, via a background field, introduces new physics into the system. For example, a modulated background field, induced by dispersive waves \cite{Coen_Optica2017,Matsko_EPJD2017} or through electro-optical modulation \cite{Coen_NC2015Tweezing}, introduces a trapping potential and traps the relative positions of solitons and the background field. Moreover, a modulated background can enable the formation of regular arrays called soliton crystals \cite{Papp_NP2017Crystal}.

In all of these cases, however, the solitons are co-propagating and the relative positions of interacting solitons will be fixed. Here, we consider a trapping mechanism that occurs between coherently pumped counter-propagating (CP) solitons \cite{Vahala_NP2017Counter,Gaeta_OL2018CP}. It is shown experimentally and theoretically that a new manifestation of soliton trapping arises for non-degenerately pumped CP solitons in the presence of optical backscattering.  Specifically, the solitons are trapped on average, but exhibit a periodic relative motion at the pump detuning frequency, $\delta \nu_\text{P}$.  This forced oscillatory motion is observed after the CP solitons are coupled outside the cavity and features an amplitude that is larger than the soliton pulse width. It also introduces an observable fine structure into the comb lines in the form of sidebands separated by $\delta \nu_\text{P}$.

\begin{figure*}[t!]
\begin{centering}
\includegraphics[width=\linewidth]{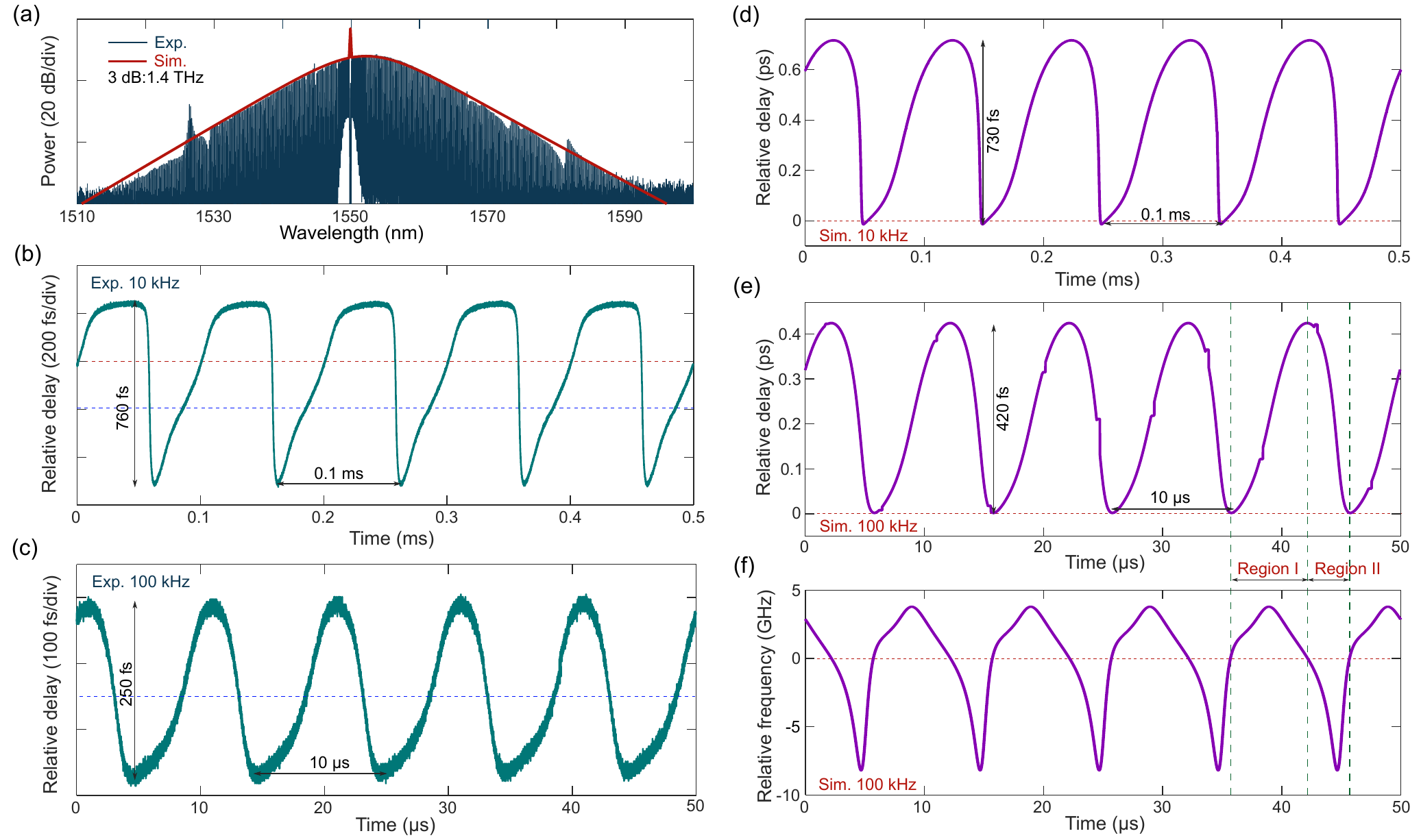}
\captionsetup{singlelinecheck=no, justification = RaggedRight}
\caption{{\bf Observation of oscillatory motion in CP soliton trapping.} (a) Optical spectrum of the CP solitons with a 3 dB bandwidth of 1.4 THz. The red line is the simulated spectral envelope and is in agreement with experimental measurement. (b, c) BOC measured relative soliton temporal motion when the pump frequency detuning is 10 kHz and 100 kHz. The motion frequency is measured to be equal to the counter-pumping frequency detuning. The red and blue dashed lines indicate the center of motion for 10 kHz and 100 kHz detuning, respectively; and they are shifted by about 200 fs as shown in panel b. (d, e) Simulations of relative soliton temporal motion when the two pumps are detuned by 10 kHz and 100 kHz, respectively. The red dashed lines indicate the zero delay. (f) Numerically calculated relative spectral center frequency between two CP solitons for $\delta\nu_\text{P}$=100 kHz, showing periodic variation. Depending upon the sign of the relative frequency, the motion can be separated into two regions as indicated by the green dashed lines.}
\label{Fig2Motion}
\end{centering}
\end{figure*}

\begin{figure}[t!]
\begin{centering}
\includegraphics[width=\linewidth]{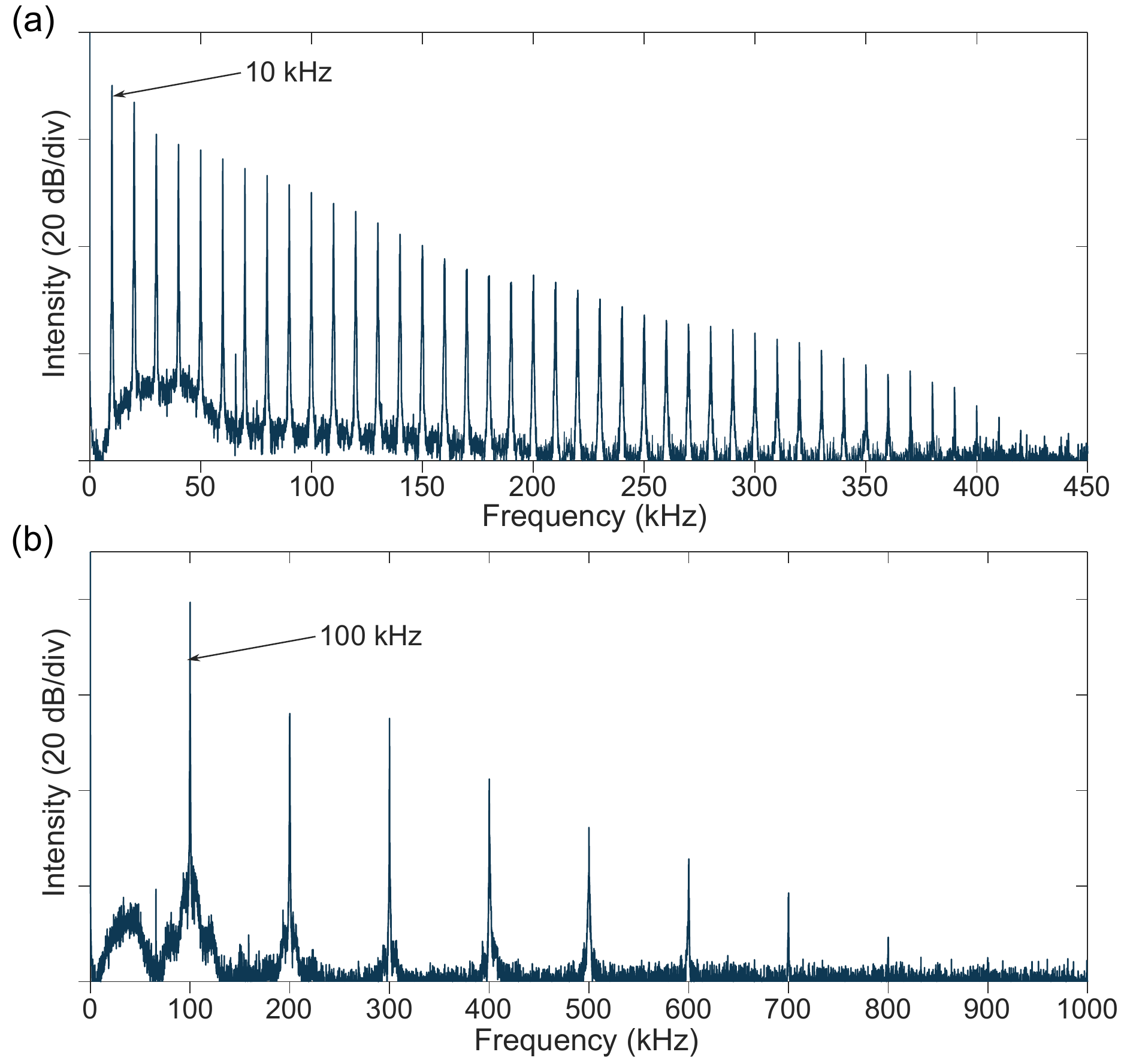}
\captionsetup{singlelinecheck=no, justification = RaggedRight}
\caption{{\bf Measured electrical spectra of the beat between the CP solitons.} (a) Electrical spectrum when beating two CP soltions with $\delta \nu_\text{P}$ = 10 kHz. (b) Electrical spectrum when beating two CP soltions with $\delta \nu_\text{P}$ = 100 kHz.}
\label{Fig3Sidebands}
\end{centering}
\end{figure}


Backscattering within a microresonator allows the fields of both the CP solitons and their respective pumps to couple into the opposing propagation directions of the resonator \cite{Vahala_NP2017Counter}. This provides a way for each soliton to interact with a co-propagating replica of the other soliton. Consider first the case of degenerate pumping frequencies. Here, a soliton propagating along say the clockwise (CW) direction will be able to interact with the replica counter-clockwise (CCW) soliton. The CCW and CW solitons share the same longitudincal mode family and similar pump conditions (same pump frequency and similar pump power). Moreover, a feature of coherent pumping is that the pump frequency coincides with a soliton comb frequency. For these reasons both the envelope and the carrier of the CP solitons will have closely matched velocities. Adding to this situation, the Kerr effect mediated interaction allows the two solitons to readily trap one another via the backscattered replica fields. This occurs in a way very similar to trapping of conventional co-propagating solitons \cite{Menyuk_OL1987,Cundiff_PRL1999,Gordon_OL1989}. Thus, even while the solitons are propagating in opposite directions, their group and phase velocities become locked. A caveat here concerns the nature of the replica solitons. Generally, backscattering would not be expected to occur at a single point, but instead to exhibit a complex spatial profile. As a result, the trapping itself would be expected to reflect the complexities of this scattering process. Indeed, evidence of cases where counter-propagating solitons are only weakly locked (i.e., spatially broad replica solitons) are observable in the measurements. 


Next consider the impact of detuning the CW and CCW pumping frequencies on this process. When the pump detuning frequency ($\delta \nu_\text{P}$) is small in comparison to the cavity free-spectral-range (FSR),  the CP solitons are still bound to each other. However, on account of the coherent nature of the pumping, the frequency of each soliton comb line will be shifted by $\delta \nu_\text{P}$ relative to the replica of the other soliton comb.  This frequency shift is set by the external CW and CCW pumps and cannot be pulled towards zero as in conventional soliton trapping. In other words, the soliton carriers and phase velocities will not be synchronized. Thus, each of the backscattered CW (CCW) comb lines can be regarded as a sideband that modulates the CCW (CW) comb lines. As shown below, this causes each soliton to experience a periodically time-varying spectral center frequency change, $\delta \nu_s$.  And, in turn, their group velocity will vary with the spectral center as $\delta (1/v_g)=2\pi\beta_2 \delta \nu_s$, where $\beta_2$ is the group velocity dispersion (negative for cavities that support solitons) \cite{Gordon&Haus_OL1986}. This velocity modulation causes periodic relative motion of the CP solitons with a period of 1/$\delta \nu_\text{P}$ within their traps.  Finally, since the soliton group velocity and round-trip time are modulated periodically with a considerable amplitude, sidebands  (spacing  $\delta \nu_\text{P}$) emerge around the main comb lines (Fig. \ref{Fig1Picture}(b)).

Observation of the relative motion requires high temporal resolution. We use a balanced optical cross correlator (BOC) to record this motion (experimental setup shown in Fig. \ref{Fig1Picture}(c)). The BOC converts temporal motion into a voltage signal with a steep discriminator slope \cite{Kartner_LPR2008}. A single laser pump is used and distinct pumping frequencies for CW and CCW directions are produced by two acousto-optical modulators (AOMs).  The CP solitons are generated in a 22 GHz high-Q silica wedge micoresonator \cite{Vahala_NP2012Wedge,Vahala_Optica2015,Vahala_NP2017Counter}. The corresponding optical spectrum of one of the solitons is shown in Fig. \ref{Fig2Motion}(a) and has a 3 dB bandwidth of 1.4 THz so that the soliton duration is deduced to be 125 fs (equivalently, 220 fs for full-width-half-maximum, FWHM, pulse width). The soliton streams (with pumps suppressed) are dispersion compensated by pulse shapers \cite{Weiner_RSI2000} and amplified to feed into the BOC.

The output of the BOC for pump detuning frequencies of 10 kHz or 100 kHz is shown in Figs. \ref{Fig2Motion}(b), (c). The ability to reliably observe a signal implies that the repetition rates of the two solitons are locked on average, since otherwise the two inputs would temporally walk-off due to non-synchronized repetition rates \cite{Kartner_LPR2008}. 
The peak-to-peak oscillation amplitude is observed to be as large as 0.8 ps which is more than twice the soliton FWHM pulse width. This is also nearly 2\% of the round-trip time (46 ps). Furthermore, the motion is not sinusoidal but is asymmetric (sawtooth-like for the 10 kHz detuning case). Significantly, the oscillation frequency is equal to the pump frequency detuning ($\delta\nu_\text{P}$). The center of the relative motion trajectory is also plotted in Figs. \ref{Fig2Motion}(b), (c). It suggests that the CP solitons will oscillate around different centers in the trap when the pump detuning frequency varies. This oscillatory motion was also observed to exist for small detunings, $\delta\nu_\text{P}<$1 Hz.

To further confirm the experimental observations, numerical simulations based on coupled generalized Lugiato-Lefever equations (LLEs) were performed  \cite{Vahala_NP2017Counter,Coen_OL2013,Chembo_PRA2013}. Details on the simulation are provided in the Supplementary Materials. The simulated soliton spectrum is in excellent agreement with experiments (see Fig. \ref{Fig2Motion}(a)). Moreover, upon detuning the counter-pumping frequencies, the CP solitons undergo periodic relative motion in simulations. Representative plots of the relative soliton motion are shown in Figs. \ref{Fig2Motion}(d), (e). The motional frequency equals $\delta \nu_\text{P}$, and both the trajectory and amplitude are reasonably consistent with experimental measurements. For example, the sawtooth-like behavior is numerically reproduced (Fig. \ref{Fig2Motion}(d)).  

To test the hypothesis that the relative motion is driven by the detuned-pump-induced soliton spectral-center-shift, we numerically calculated the relative spectral center frequency between the two solitons Fig. \ref{Fig2Motion}(f). It exhibits periodic oscillation around 0 Hz. The positive and negative relative frequency regions correspond to forced motion where the derivative of the relative delay  (i.e., relative group velocity) is positive or negative, respectively (see the green dashed vertical lines in Fig. \ref{Fig2Motion}(e), (f)). Accordingly, the experimental and numerical observations validate the existence of the coherent-pump forced oscillatory motion between CP solitons in the presence of backscattering.

Finally, we experimentally verify that the forced motion introduces fine structure sidebands into the comb lines. For this measurement,  the two CP soliton microcombs are heterodyned on a balanced photodetector. The recorded electrical spectra, shown in Fig. \ref{Fig3Sidebands},  contain multiple RF tones for both pump detuning of 10 kHz and 100 kHz. The lack of any tone other than those at integer multiples of  $\delta \nu_\text{P}$ shows that only fine structure sidebands spaced by $\delta \nu_\text{P}$ are present in the optical comb spectra. 


The range of detuning frequencies over which this trapping occurs depends upon the backscattering level. A prior study of CP solitons shows that beyond pump detuning frequencies of several 100 kHz, the CP solitons unlock and feature independently controllable repetition rates \cite{Vahala_NP2017Counter}. For small $\nu_\text{P}$, a locking zone with the same repetition rate for CP solions was observed in that work, which was not explored there but can be understood to result from the average locking described here. In the regime of independent repetition rates control, CP solitons can also exhibit interactions via backscattering which can cause the solitons to experience optical phase locking at specific pump detuning frequencies (e.g., several MHz) \cite{Vahala_NP2017Counter}.

In summary, we observe a new manifestation of soliton trapping in high-Q microresonators as a result of the coherently pumped nature of the solitons. This phenomenon results from slightly detuned counter-pumping of CP solitons in the presence of optical backscattering. The detuned counter pumps cause a periodic variation of both the spectral center frequencies of the CP solitons and their relative positions. Our measurements show that this trapping does not require the solitons to be constrained within a small temporal range.  The oscillatory motion also inserts fine structure sidebands into the soliton microcomb spectrum that may affect some comb applications. This mechanism may be engineered to program and control the pulse timing of microresonator solitons and could also be important for the development of microcomb-based gyroscopes.

\vspace{3 mm}

This work is supported by the Air Force Office of Scientific Research (FA9550-18-1-0353) and the Kavli Nanoscience Institute. C.B. gratefully acknowledges a postdoctoral fellowship from the Resnick Institute at Caltech.

\bibliography{main}
\end{document}